\documentclass[manuscript]{emulateapj}

\slugcomment{}

\shorttitle{Ultraviolet Color-Magnitude Relation of Virgo Early-type Dwarf Galaxies}

\shortauthors{Kim et al.}

\begin{document}

\title{Color-Magnitude Relations of Early-type Dwarf Galaxies in the Virgo Cluster: An Ultraviolet Perspective}

\author{Suk Kim\altaffilmark{1}, Soo-Chang Rey\altaffilmark{1, 2}, Thorsten Lisker\altaffilmark{3}, and Sangmo Tony Sohn\altaffilmark{4}}
\altaffiltext{1}{Department of Astronomy and Space Science, Chungnam National University, Daejeon 305-764, Korea; screy@cnu.ac.kr}
\altaffiltext{2}{Department of Physics and Astronomy, Johns Hopkins University, 3400 North Charles Street, Baltimore, MD 21218, USA}
\altaffiltext{3}{Astronomisches Rechen-Institut, Zentrum f\"ur Astronomie der Universit\"at Heidelberg (ZAH), M\"onchhofstra\ss e 12-14, D-69120 Heidelberg, Germany}
\altaffiltext{4}{Space Telescope Science Institute, 3700 San Martin Drive, Baltimore, MD 21218, USA}
\begin{abstract}
We present ultraviolet (UV) color-magnitude relations (CMRs) of early-type dwarf galaxies in the Virgo cluster, based on Galaxy Evolution Explorer (GALEX) UV and Sloan Digital Sky Survey (SDSS) optical imaging data. We find that dwarf lenticular galaxies (dS0s), including peculiar dwarf elliptical galaxies (dEs) with disk substructures and blue centers, show a surprisingly distinct and tight locus separated from that of ordinary dEs, which is not clearly seen in previous CMRs. The dS0s in UV CMRs follow a steeper sequence than dEs and show bluer UV$-$optical color at a given magnitude.  
We also find that the UV CMRs of dEs in the outer cluster region are slightly steeper than that of their counterparts in the inner region, due to the existence of faint, blue dEs in the outer region. We explore the observed CMRs with population models of a luminosity-dependent delayed exponential star formation history. We confirm that the feature of delayed star formation of early-type dwarf galaxies in Virgo cluster is strongly correlated with their morphology and environment. The observed CMR of dS0s is well matched by models with relatively long delayed star formation. Our results suggest that dS0s are most likely transitional objects at the stage of subsequent transformation of late-type progenitors to ordinary red dEs in the cluster environment. In any case, UV photometry provides a powerful tool to disentangle the diverse subpopulations of early-type dwarf galaxies and uncover their evolutionary histories.
\end{abstract}

\keywords{galaxies: clusters: individual (Virgo) --- galaxies: dwarf --- galaxies: star formation --- ultraviolet: galaxies}

\section{Introduction}
It is well established that early-type galaxies in clusters form a tight and well-defined color-magnitude relation (CMR) in the optical bands (Visvanathan \&\ Sandage 1977; Sandage \&\ Visvanathan 1978a, b) such that brighter galaxies are generally redder. The physical origin (e.g., metallicity vs. age) of the CMR is still a matter of debate. Nonetheless, it is believed that the CMR is an important tool for understanding star formation histories (SFHs) of early-type galaxies and their links to the galaxy formation scenarios.  

Since the ultraviolet (UV) flux of an integrated population is a good tracer of recent star formation activities, the CMR in the UV band provides an important constraint on SFH in galaxies. In recent years, UV CMRs for early-type galaxies in the local Universe based on Galaxy Evolution Explorer (GALEX) observations have shown a substantially larger scatter in the bluer UV$-$optical colors than optical CMRs. This indicates that a large fraction (10 $\sim$ 30\% of the early-type galaxies examined with GALEX) of early-type galaxies have experienced low level ($\lesssim$ 1 M$_\odot$yr$^{-1}$) residual star formation over the last 8 billion years (Yi et al. 2005; Kaviraj et al. 2007, 2008; Schawinski et al. 2007; Haines et al. 2008).

The SFHs of galaxies are strongly dependent on their masses (Ferreras \&\ Silk 2000; Caldwell et al. 2003; Nelan et al. 2005; Thomas et al. 2005) and low-mass dwarf galaxies have much more extended SFHs with longer timescales of gas consumption for star formation compared to the massive galaxies (e.g., Mateo 1998; Grebel \&\ Gallagher 2004 for Local Group and Gavazzi et al. 2002 for Virgo cluster). Moreover, low-mass galaxies are more susceptible to external processes such as galaxy harassment, tidal interaction, and ram-pressure stripping that ultimately affect their star formation and evolutionary histories. In this context, it is important to compare the UV CMR of early-type dwarf galaxies with that of the massive counterparts. 

Boselli et al. (2005) first studied the UV properties of early-type galaxies in the Virgo cluster using GALEX data. They found that UV CMRs show a discontinuity between massive and dwarf early-type galaxies in contrast to what is observed at optical wavelengths. They suggested that the residual star formation activity is more important in low mass early-type dwarf galaxies, while the UV flux in the massive counterparts is dominated by hot, evolved old stellar populations.  However, their results are based on a sample restricted to bright dwarf galaxies detected from GALEX observations of its early operation. In order to understand the SFH in galaxies related to their masses using UV CMRs, a larger sample is required to reach into the fainter and lower mass dwarf regime.

Recently, several subclasses of dwarf ellipticals (dEs) with morphological substructures such as disks, spiral arms, bars, and blue centers, have been discovered (Jerjen, Kalnajs, \&\ Binggeli 2000; Barazza, Binggeli, \&\ Jerjen 2002; Lisker et al. 2006a, b). These galaxies constitute a significant fraction of bright dEs in the Virgo cluster (Lisker et al. 2007). Their properties might be related to recent or ongoing star formation activities and distinct environmental effects. The discovery of these objects hints that the early-type dwarf galaxies are heterogeneous objects, which originated from various channels of evolutionary scenarios (see Lisker 2009 for a review). Here, we present new UV CMRs of early-type dwarf galaxies in the Virgo cluster using extensive GALEX UV photometric data in combination with SDSS data. Our goal is to study whether the various subclasses of early-type dwarf galaxies show different sequences in UV CMRs related to their star formation and evolutionary histories.

\section{Data and Analysis}
We used UV images from the GALEX Release 3 (GR3) dataset. GALEX observed the Virgo cluster as part of the All-sky Imaging Survey (AIS), Nearby Galaxy Survey (NGS), and Deep Imaging Survey (DIS) in two UV bands : far-ultraviolet (FUV; 1350$-$1750\AA) and near-ultraviolet (NUV; 1750$-$2750\AA).  GALEX imaged 97 fields of Virgo cluster covering a total $\sim$ 82 deg$^2$.  The depth of each field varies in accordance with its survey mode: 16 NGS fields (NUV$\sim$3,000s, FUV$\sim$1,500s), 80 AIS fields (NUV, FUV$\sim$100s), and 1 DIS field (NUV$\sim$22,000s). Most NGS fields (12 of 16) cover the regions within angular distance of 2 degree from the M87. Using SExtractor (Bertin \&\ Arnouts 1996), we performed photometry for all detected objects. For this, we required fluxes at least 1$\sigma$ above the sky noise. We adopted MAG$\_$AUTO(total) as the source magnitude. Flux calibrations were applied to bring the final photometry into the AB magnitude system (Oke 1990). The typical errors are 0.10 mag and 0.14 mag in the NUV and FUV, respectively.

We take advantage of the comprehensive sample of certain or possible cluster members classified as dE and dwarf lenticular (dS0) galaxies in the Virgo Cluster Catalog (VCC) of Binggeli et al. (1985). In addition, we include 11 lenticular galaxies (S0s) of the VCC with optical magnitudes similar to dS0. The cross-identification between 774 VCC early-type dwarf galaxies and GALEX photometry results in 193 and 59 galaxies in the NUV and FUV band, respectively. All matched objects were visually inspected and we retained objects with clear detection. NGS fields reach limiting magnitudes of $\sim$ 23.0 mag in the NUV and FUV, while AIS ones reach $\sim$ 22.0 mag. All FUV-detected galaxies are also detected in the NUV. The resulting sample includes fainter dwarf galaxies compared to previous UV studies using the GALEX Internal Release (IR1.0) (Boselli et al. 2005). We secured galaxies down to m$_{B}$ $\sim$ 20 mag. GALEX UV data have been combined with SDSS r-band data from SDSS Data Release 5 (DR5). The SDSS photometric pipeline fails to measure accurately the local sky flux around Virgo dEs and thus the total magnitude (see Lisker et al. 2007 for the details). Therefore, we performed our own sky subtraction and photometric measurement, following the procedure of Lisker et al. (2007). Only foreground Galactic extinction correction for each galaxy is applied (Schlegel et al. 1998). We use the reddening law of Cardelli et al. (1989) to derive the following : R$_{NUV}$=8.90, R$_{FUV}$=8.16, and R$_{r}$=2.72 . We adopt a Virgo cluster distance of 15.9 Mpc, i.e.\ a distance modulus m$-$M=31.01 mag (Graham et al. 1999).

\section{Results}
\subsection{Ultraviolet Color-Magnitude Relations of Early-Type Dwarf Galaxies}

In Figure 1, we present optical (Fig. 1a), NUV (Fig. 1b), and FUV (Fig 1c) CMRs for dEs (red circles) and dS0s (yellow circles).
Of galaxies classified as dS0s, a substantial fraction corresponds to dEs with disk substructures (stars, Lisker et al. 2006a) or blue centers (triangles; Lisker et al. 2006b). Note that, in what follows, we refer to dS0s and peculiar dEs (disk substructure or 
blue center) collectively as dS0s. We also include blue compact dwarf galaxies (BCDs, squares) drawn from the VCC, for comparison purposes. UV CMRs follow the general trend of the optical CMR, i.e., early-type dwarf galaxies become progressively bluer with decreasing optical luminosity. However, the UV colors span a much wider range than the optical CMR, owing to the wide baseline of UV to optical colors: while $g-r$ only spans a range of $\sim$0.6 mag, NUV$-r$ and FUV$-r$ varies up to 4.5 mag and 6.0 mag, respectively.

The most interesting feature in our UV CMRs is that dS0s form a tight sequence which is clearly distinct from that of normal dEs. In UV CMRs, dS0s follow a steeper sequence than dEs (dotted line in Fig. 1a-c gives the mean of dS0s).  Meanwhile, the optical CMR of dS0s (see Fig. 1a) is not much different from that of normal dEs. We note that Boselli et al. (2005, 2008) were not able to observe such features in their UV CMRs mainly due to their limited sample of dwarf galaxies. In addition, the faint end of the dS0 sequence in UV CMRs appears to be linked to the BCDs. Note that some galaxies originally classified as BCDs in the VCC have a similar appearance like dEs with blue centers (Lisker et al. 2006b): the visual classification between dE with blue center and BCD appears to have a smooth transition. This is now confirmed by our UV CMRs. Furthermore, several studies claimed that BCDs might be potential progenitors of dEs (see Lisker 2009 and references therein). Our UV CMRs shown in Fig. 1 indicate that dS0s evidently have different stellar population properties as compared to normal dEs.

Since the UV flux is sensitive to young ($\lesssim$ 1 Gyr) stellar populations, the bluer UV colors of dS0s at fixed luminosity implies that dS0s have experienced recent or ongoing star formation activities whereas dEs have been relatively quiescent in the past few Gyrs. To confirm this, we examined SDSS spectra and available literature (Boselli et al. 2008; Michielsen et al. 2008; Paudel et al. 2010) and found that the majority of dS0s show relatively strong H$\alpha$ emission and/or H$\beta$ absorption lines. Interestingly, dS0s showing H$\alpha$ emission lines are systematically less luminous and have strong NUV and FUV fluxes. We found that 85\%\ of 13 faint ($M_{r}$ $>$ -16.9) dS0s exhibit H$\alpha$ emission lines with EW $>$ 2 \AA. Meanwhile, dS0s that show strong H$\beta$ absorptions are preferentially located in the luminous part of the sequence (63\%\ of 16 dS0s with $M_{r}$ $<$ -16.9 show H$\beta$ EW $>$ 2.5 \AA). On the other hand, 6\% and 36\% of normal dEs show hints of ongoing and post star formation, respectively, according to the SDSS spectra of 83 sample with $M_{r}<$-13.7. Therefore, spectroscopic results confirm the systematically distinct sequence of dS0s in UV CMRs, implying evidence of ongoing or post star formation (see also Boselli et al. 2008).

The environment in a cluster plays an important role in the formation and evolution of the member galaxies. Our large sample size allows us to examine this effect for early-type dwarf galaxies in the Virgo cluster. In order to investigate the environmental dependence of the CMRs, we divide the sample into two based on their distance from the giant elliptical galaxy M87. Given that dwarf galaxies with angular distances of less than 2 degree from the center of M87 are dynamically connected with M87 (Binggeli et al. 1987), we adopt this angular distance for dividing our sample into inner and outer dwarf galaxies. While handful outer dwarf galaxies (three in the NUV) might be associated with other giant galaxy, M49, those galaxies do not change our results. In Fig. 1d and 1e, we present the spatial distribution of early-type dwarf galaxies detected in the GALEX NUV and FUV, respectively. In all panels of Fig. 1, large and small symbols denote dwarf galaxies in the inner and outer region, respectively. As suggested by Lisker et al. (2007), dS0s are not centrally clustered around M87. 

We have fitted CMRs of dEs in two different regions using a first-order least squares method. In the optical band (Fig. 1a), CMRs for dEs in the inner and outer regions are nearly identical. However, in the UV (Fig. 1b, c), dEs in the outer region (dashed line) have a steeper relation than those in the inner region (solid line). In Fig. 1a-c, we also plot color distributions of dEs in the inner (solid  histogram) and outer (dotted histogram) region. Again, $g-r$ color distributions between inner and outer dEs are indistinguishable (histograms in Fig. 1a). Yet, the UV$-$optical color distributions of dEs in two different regions are very different; while inner dEs are confined to the redder side (NUV$-r$ $\gtrsim$ 3.5, FUV$-r$ $\gtrsim$ 5), those in the outer region exhibit wider distributions extending to bluer colors (histograms in Fig 1b, c). This mainly results from the contribution of faint ( Mr $\gtrsim$ -14) dEs in the outer region showing bluer UV$-$optical colors. Our results are in good agreement with those from other studies where early-type dwarf galaxies with hints of star formation are found to be preferentially located in the outskirts of clusters (Drinkwater et al. 2001; Conselice et al. 2003; Smith et al. 2008, 2009). In Table 1, we present the main relations of CMRs for early-type dwarf galaxies.

\subsection{Comparison with Population Models}
We now compare observed optical and UV CMRs with evolutionary stellar population models. The models of Bruzual \&\ Charlot (2003) are used for the young ($<$ 1 Gyr) stellar populations. We combine them with the models of Yi (2003) in order to represent flux of old ($>$ 1 Gyr) stellar populations. We assume a delayed exponential SFH given by

\[SFR(T,\tau)=(T/\tau^2)\times exp(-T^2/2\tau^2),\]

where T is a time from the onset of star formation and $\tau$ is the star formation timescale regulating the delay of the maximum star formation rate (SFR) and the steepness of its decay (Gavazzi et al. 2002). 

In Figure 2, we show population models overlaid on the observed CMRs. Model lines are computed for an epoch of T = 13 Gyr for three cases: dEs in the outer region (left columns), dEs in the inner region (middle columns), and dS0s (right columns). For each case, we adopt different ranges of $\tau$ that best match the observed CMRs. The $\tau$ values increase with decreasing luminosity of the galaxy. The different model lines (dotted lines) in each panel refer to those obtained for six different metallicities (Z = 0.0001, 0.0004, 0.001, 0.004, 0.01, 0.02 from bottom to top), which cover the observed metallicity range of dwarf galaxies (e.g., Barazza \&\ Binggeli 2002; Jerjen, Binggeli, \&\ Barazza 2004). Three large rectangles in each panel denote the model predictions adopting an empirical luminosity-metallicity relation for dEs (Barazza \&\ Binggeli 2002). Since the UV flux is very sensitive to even small variations on the SFR, NUV$-r$ and FUV$-r$ colors show higher dependence on $\tau$ than $g-r$ color.

The observed CMRs of dEs in the inner region (Fig. 2a-c) are well matched by the model lines with small $\tau$ range (2 $<$ $\tau$ $<$ 3.5 Gyr). Since a small $\tau$ range implies a burst of star formation with a short timescale at an early epoch, the resulting CMR is expected be relatively flat. This is consistent with our observed CMRs as shown in Fig. 2a-c. 

Unlike dEs in the inner region, dEs in the outer region (Fig. 2d-f) extend to fainter ($M_{r}$ $\gtrsim$ -14) magnitudes in the CMRs. These faint dEs show on average predominantly bluer UV$-r$ colors than the extrapolated mean line of dEs in the inner region (see Fig. 1). That is, fainter dEs become bluer more rapidly than bright dEs do with decreasing magnitude. Overall, this makes the CMRs of dEs in the outer region different from that of the inner region dEs. Such environmental dependences of CMRs were also observed in other studies (Tanaka et al. 2005; Baldry et al. 2006; Haines et al. 2007; Gavazzi et al. 2010). In both optical and UV CMRs, the distribution of luminous ($M_{r}$ $\lesssim$ -14) dEs in the outer region is similar to that of the counterparts in the inner region, and are well matched by models with small $\tau$ range (2 $<$ $\tau$ $<$ 3.5 Gyr, two solid lines for Z=0.0001 and 0.02 in Fig. 2d-f). On the other hand, fainter dEs in the outer region lie significantly to the bluer side of these model predictions. In order to match the distribution of these faint and blue dEs, models with wider $\tau$ range (2 $<$ $\tau$ $<$ 6 Gyr, dotted lines) are required, which translates into a relatively higher SFR at the current epoch, owing to a delayed star formation. 

As for the dS0s, models with large $\tau$ range (2 $<$ $\tau$ $<$ 7 Gyr) are good matches to the observed CMRs (Fig. 2g-i), describing well the steep sequence of dS0s. Therefore, we conclude that dS0s have likely experienced relatively long delayed star formations similar to dEs found in the outer region. This is consistent with observational results indicating residual or ongoing star formation activities (Lisker et al. 2006a, b, 2007; Boselli et al. 2008).

\section{Discussion}

The dS0s seem to share properties with both normal dEs and late-type dwarf galaxies. While their overall appearance is closer to that of normal dEs, their various characteristics are distinct in many aspects. First, in the optical images, they exhibit substructures such as disks, blue centers, or irregular central features (Lisker et al. 2006a, b, 2007). Second, a number of them show Balmer emissions and/or absorptions as well as HI gas content (Lisker et al. 2006a, b, 2007, 2008; Boselli et al. 2008). Third, they are not clustered around the center of the potential well and have an asymmetric velocity distribution with multiple peaks, indicative of an unrelaxed population from recent infall (Lisker et al. 2007). 

If we assume that a fraction of early-type dwarf galaxies originated from galaxies that fell into the cluster's potential well, one could imagine that a galaxy in the denser regions of the cluster would have experienced an earlier destruction of substructures, accompanied with rapid truncation of its star formation, due to strong environmental effects (e.g., galaxy harassment, Moore et al. 1999 and ram-pressure stripping, Gunn \&\ Gott 1972, see Boselli \&\ Gavazzi 2006; Boselli et al. 2008; Smith et al. 2010 for the details). Our UV CMR results support this view, in the sense that bluer UV$-$optical colors of dS0s are interpreted as contribution from young stellar populations. Of course, this assumes that the progenitor galaxies had experienced star formation activities before the infall into the cluster's potential well. The hypothesis is also consistent with recent results that rotationally supported dEs with disk features are on average younger than pressure supported normal dEs (e.g., Toloba et al. 2009; Michielsen et al. 2008). 
Consequently, dS0s are most likely subclasses of transitional objects between red dEs and late-type bluer dwarf galaxies, and will eventually migrate to red-sequence early-type dwarf galaxies (Boselli et al. 2008; Haines et al. 2008; Gavazzi et al. 2010).

In this regard, the study of the diverse subpopulations of early-type dwarf galaxies might play an important role in the issue of growth of the low-mass part of the red sequence via the evolution of the blue sequence and green valley, under the downsizing paradigm (Cowie et al. 1996) at the current epoch. A comparative study of the UV CMRs for dwarf galaxies in various clusters and groups with different densities and dynamical conditions will provide additional insight into the physical processes of galaxy transformation via cluster environmental effects. We defer presenting our results on such topics to our forthcoming papers.

\acknowledgments
We are grateful for the clarifications and improvements suggested by an 
anonymous referee.
This research was supported by Basic Science Research Program through the
National Research Foundation of Korea (NRF) funded by the Ministry of
Education, Science and Technology (No. 2009-0070263)
and the NRF grant funded by the Korea government (MEST)
(No. 2009-0062863).
T.L.\ is supported within the framework of the Excellence Initiative
by the German Research Foundation (DFG) through the Heidelberg
Graduate School of Fundamental Physics (grant number GSC 129/1).

{\it Facilities:} \facility{GALEX}

\clearpage

\begin{figure}
\epsscale{.80}
\plotone{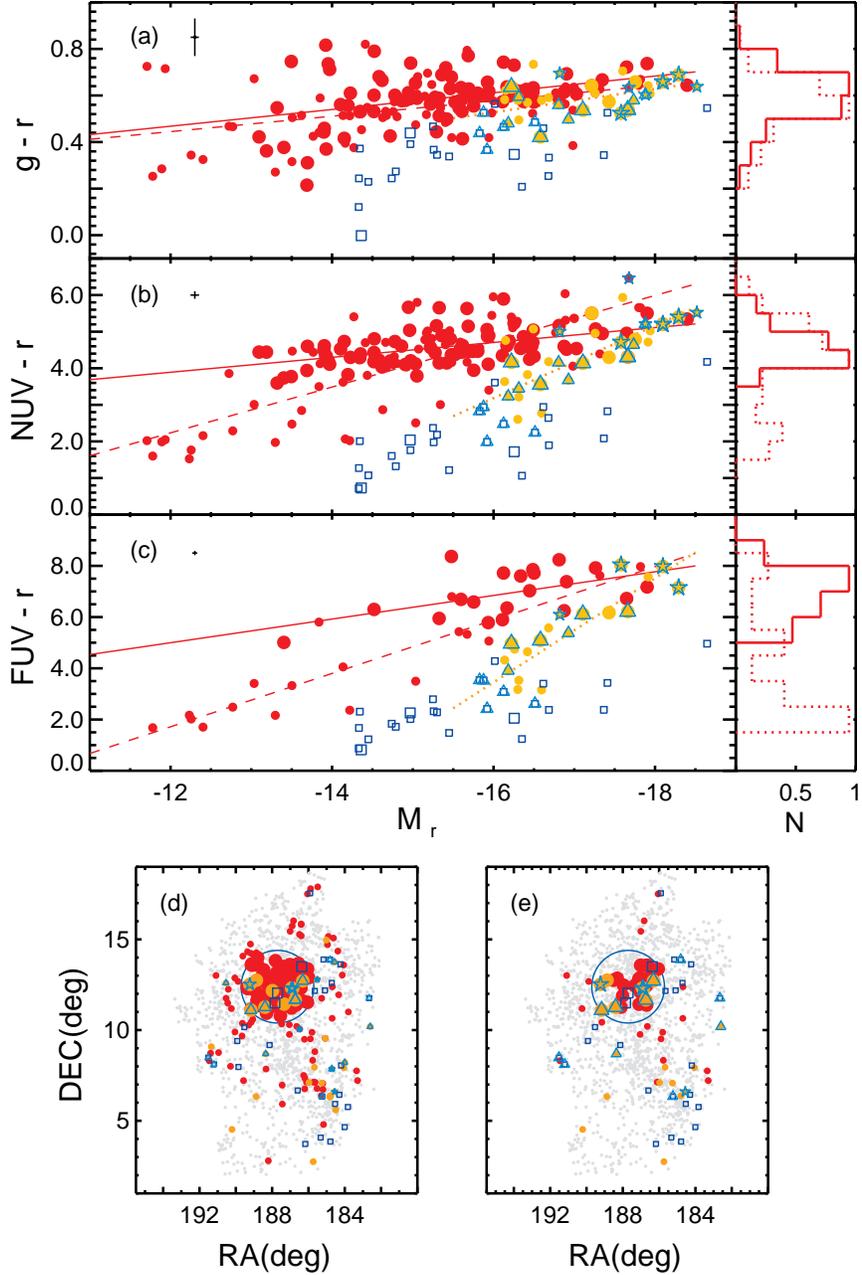}
{~~~~~}
\caption{Color-magnitude relation (CMR) and spatial distribution of early-type dwarf galaxies. (a) Optical CMR for various early-type dwarf galaxies; dEs (red filled circles), dS0s (yellow filled circles), dEs with disk (stars), dEs with blue center (triangles). Additionally, we plot blue compact dwarf galaxies as squares. We divide the dwarf galaxies into two subsamples according to their spatial distribution; inner region (large symbols) and outer region (small symbols). The solid and dashed line represents the linear least squares fit to the observed means for dEs in the inner and outer region, respectively, while dotted line is for dS0s. Mean errors for the sample are shown as error bars in the upper left corner. Right to the CMR, we present color distributions of dEs in the inner (solid histogram) and outer (dotted histogram) region normalized to the total number of sample. (b, c) Same as (a), but for UV CMRs of galaxies detected in the NUV and FUV, respectively. (d, e) Spatial distribution of galaxies detected in the NUV and FUV, respectively. Large circle indicates boundary of 2 deg from the M87 for dividing galaxies in the inner and outer region.\label{fig1}}
\end{figure}

\begin{figure}
\epsscale{.90}
\plotone{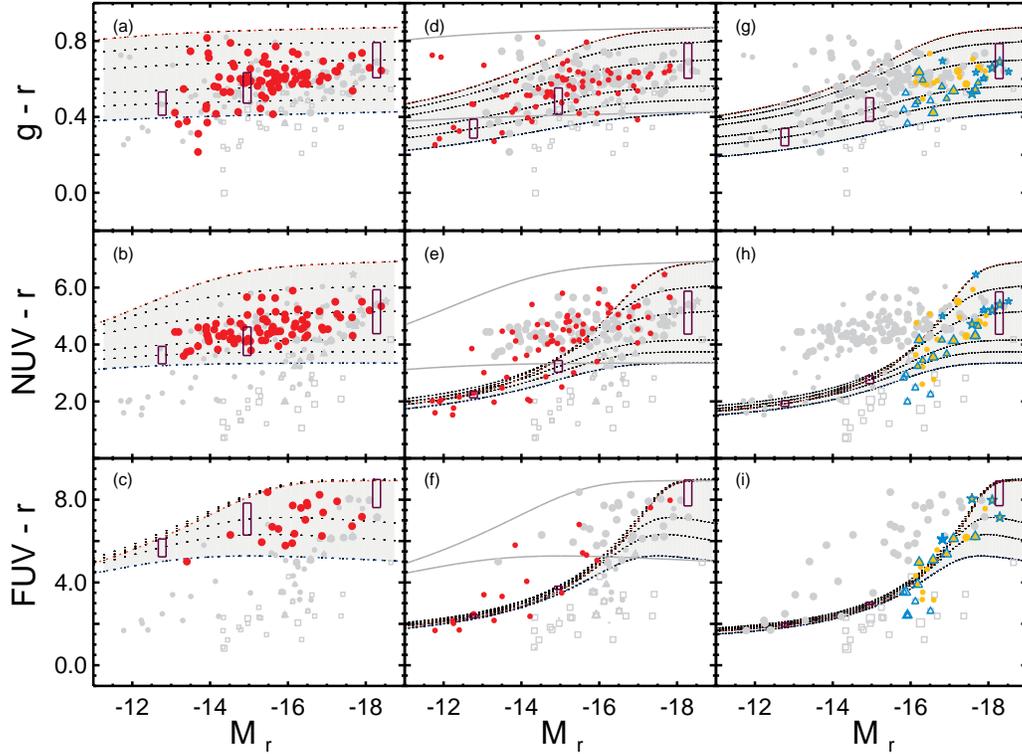}
\caption{Observed color-magnitude relations in comparison with population models for three cases of galaxies; dEs in the inner region (left column), dEs in the outer region (middle column), and dS0s (right column). The model lines are produced by assuming a delayed exponential star formation history. For each case, we adopt different relations of the characteristic star formation timescale $\tau$ (Gyr) with magnitude, to match the sequence of observed CMRs (see main text for the details): $\tau$ = 0.19M$_{r}+$5.56 for the inner region dEs, $\tau$ = 0.50M$_{r}+$11.50 for the outer region dEs, and $\tau$ = 0.63M$_{r}+$13.88 for dS0s. Different model lines in each panel are those with different metallicities of Z = 0.0001, 0.0004, 0.001, 0.004, 0.01, 0.02 (from bottom to top). In each panel, the gray symbols in the background represent the whole sample. The symbols of the galaxies in each case are same as Fig. 1. Three large rectangles in each panel denote the model predictions adopting empirical luminosity-metallicity relation and mean error of [Fe/H] at given magnitudes for dEs (Barazza \&\ Binggeli 2002).\label{fig2}}
\end{figure}

\begin{table}
\label{tablefit}
\caption {Relations for early-type dwarf galaxies}
\[
\begin{array}{ccrrrr}
\hline
\noalign{\smallskip}
x  &y  & a~~~~~~ & b~~~~~~~& R~~  & rms \\

\noalign{\smallskip}
\hline
\hline
\noalign{\smallskip}
 \multicolumn{6}{c}{$Dwarf Ellipticals in the Inner Region$}   \\
\noalign{\smallskip}
\hline
\noalign{\smallskip}
M_r	&   g-r      &~~~ -0.04\pm0.01 &    0.04\pm0.14     &-0.40 & 0.10  \\
M_r	&   NUV-r    &~~~ -0.20\pm0.04 &    1.45\pm0.58     &-0.52 & 0.41  \\
M_r	&   FUV-r    &~~~ -0.46\pm0.18 &  - 0.54\pm2.90     &-0.52 & 0.76  \\

\hline
\noalign{\smallskip}
\multicolumn{6}{c}{$Dwarf Ellipticals in the Outer Region$}   \\
\noalign{\smallskip}
\hline
\noalign{\smallskip}
M_r	&   g-r      & -0.03\pm0.01 &   0.05\pm0.13        &  -0.43& 0.11  \\
M_r	&   NUV-r    & -0.63\pm0.06 &  -  5.27\pm0.88         &  -0.80 & 0.75  \\
M_r	&   FUV-r    & -1.04\pm0.13 &  - 10.79\pm1.82         &  -0.90 & 0.88  \\

\hline
\noalign{\smallskip}
 \multicolumn{6}{c}{$Dwarf Lenticulars$}  \\
\noalign{\smallskip}
\hline
\noalign{\smallskip}
M_r	&   g-r      & -0.05\pm0.02 &   -0.28\pm0.27       &  -0.47 & 0.07 \\
M_r	&   NUV-r    & -0.99\pm0.15 &  - 12.66 \pm2.52       & -0.75 & 0.67 \\
M_r	&   FUV-r    & -2.03\pm0.24 &  - 29.04\pm4.06       & -0.87 & 0.81 \\
\noalign{\smallskip}
\hline
\end{array}
\]
Note:
Cols. (1) and (2): $x$ and $y$ variables.
Cols. (3) and (4): Slope $a$ and intercept $b$ of the linear fit.
Col. (5): Pearson correlation coefficient.
Col. (6): Mean dispersion around the best fit.
\end{table}
\end{document}